\documentclass[a4paper,preprint,showpacs,amssymb]{revtex4}
\usepackage{graphicx}

\begin{document}

\title{Bulk Mediated Surface Diffusion: The Infinite System Case}
\author{Jorge A. Revelli$^{a}$, Carlos. E. Budde$^{b}$, Domingo
Prato$^{b}$ and Horacio S. Wio$^{a, c}$}
\affiliation{(a) Grupo de F\'{\i}sica Estad\'{\i}stica, Centro
At\'omico Bariloche and Instituto Balseiro, \\ 8400 San Carlos de
Bariloche, Argentina; \\ (b) Facultad de Matem\'aticas,
Astronom\'{\i}a y F\'{\i}sica, \\ Universidad Nacional de
C\'ordoba \\ 5000 C\'ordoba, Argentina; \\ (c)Departament de
F\'{\i}sica, \\ Universitat de les Illes Balears and IMEDEA \\
E-07122 Palma de Mallorca, Spain}

\begin{abstract}
An analytical soluble model based on a {\it Continuous Time Random
Walk} (CTRW) scheme for the adsorption-desorption processes at
interfaces, called bulk-mediated surface diffusion, is presented.
The time evolution of the effective probability distribution
width on  the surface is calculated and analyzed within an
anomalous diffusion framework. The asymptotic behavior for large
times shows a sub-diffusive regime for the effective surface
diffusion but, depending on the observed range of time, other
regimes may be obtained. Montecarlo simulations show excellent
agreement with analytical results. As an important byproduct of
the indicated approach, we present the evaluation of the time for
the first visit to the surface.
\end{abstract}

\maketitle

\vskip 1.truecm

\newpage

\normalsize

\section{Introduction}
\label{introduccion}

The dynamics of adsorbed molecules is a fundamental issue in
interface science \cite {v1,v4} and is crucial to a number of
emerging technologies \cite {v1,v5}. Its role is central to
phenomena as diverse as foam relaxation \cite{v6} and the
evolution of blood protein deposit \cite {v7}. Recently, the
mechanism called bulk-mediated surface diffusion has been
identified and explored. Its importance in relaxing homogeneous
surface density perturbations is experimentally well established.
This mechanism arises at interfaces separating a liquid bulk phase
and  a second phase which may be either solid, liquid, or gaseous.
Whenever the adsorbed species is soluble in the liquid bulk,
adsorption-desorption processes occur continuously. These
processes generate a surface displacement because molecules
desorb, undergo Fickian diffusion in the bulk liquid, and are then
re-adsorbed elsewhere. When this process is repeated many times,
it results in an effective diffusion of a molecule on the surface.
Bichuk and O$´$Shaughnessy \cite{Bichuk} have claimed that this
effective surface diffusion has anomalous super-diffusive
characteristics when certain range of time is considered.

Dynamical processes that display anomalous diffusion
\cite{Tsallis,Prato,Re,Zasla,Klafter,Blumen,Zumofen} have been
characterized by a non linear time dependence of the mean square
displacement of the walker for long times; that is $\langle
r^{2}(t)\rangle \;\sim\;t^{\epsilon}$ with $\epsilon\;\neq 1$
(remember that $\epsilon = 1$ corresponds to normal diffusion)
since $\langle r^2\rangle $ is the usual estimator of the square
width of the probability distribution at time $t$. Hence, for
anomalous diffusion we have that the probability distribution
width grows faster (slower) for $\epsilon > 1$ ($\epsilon <1$)
than it does for normal diffusion.

In this paper we present an analytical soluble model for the
adsorption-desorption processes based on a  {\it Continuous Time
Random Walk} (CTRW)  scheme. We calculate  the evolution with time
of the square width of the effective probability distribution on
the surface and show that, for a given range of time, this square
width growth as $t^{\epsilon}$  where $\epsilon$  depends on the
values of the adsorption and diffusion parameters.

\section{The adsorption-desorption model}
\label{modelo}

Let us start with the problem of a particle making a random walk
in the semi-infinite cubic lattice. The position of the walker is
defined by the vector $\vec{r}$ whose components are denoted by
the integer numbers $n,m,l\geq 1$ corresponding to the directions
$x$, $y$ and $z$ respectively. The displacements in the $x$ and
$y$ directions are unbounded. In the $z$ direction the particle
can move from $l=1$ to infinity.

The probability that the walker is in $(n,m,l)$ at time $t$, given that
it was at $(0,0,l_0)$ at $t_0$, $P(n,m,l;t | n,m,l_0;t_0) \equiv
P(n,m,l;t)$, satisfies the following set of coupled master
equations
\begin{eqnarray}
\label{mod1}
\dot{P}(n,m,1;t) & = & \gamma P(n,m,2;t) - \delta P(n,m,1;t),~~~~~~~~~~~~~~~~~~~~~~~~~~~~~~~~~~ \mbox{ $l = 1$} \nonumber \\
\dot{P}(n,m,2;t) & = & \alpha [P(n-1,m,2;t)+P(n+1,m,2;t)-2 P(n,m,2;t)]+  \nonumber \\
                 &   & \beta [P(n,m-1,2;t)+P(n,m+1,2;t)-2 P(n,m,2;t)]+   \nonumber \\
                 &   & \gamma P(n,m,3;t) + \delta P(n,m,1;t) - 2 \gamma P(n,m,2;t),~~~~~~~~~~~~ \mbox{ $l = 2$} \nonumber \\
\dot{P}(n,m,l;t) & = & \alpha [P(n-1,m,l;t)+P(n+1,m,l;t)-2 P(n,m,l;t)]+ \nonumber \\
                 &   & \beta [P(n,m-1,l;t)+P(n,m+1,l;t)-2 P(n,m,l;t)]+  \nonumber \\
                 &   & \gamma [P(n,m,l+1;t) +  P(n,m,l-1;t) - 2 P(n,m,l;t)],~~~~ \mbox{ $l\geq 3$}
\end{eqnarray}
where $\alpha, \beta$ and $\gamma $ are the transition
probabilities per unit time in the $x$,$y$ and $z$  directions
respectively, and $\delta$ is the desorption probability per unit
time from the boundary plane defined by $z=1$.

Taking the Fourier transform with respect to the $x$ and $y$
variables and the Laplace transform with respect to the time $t$
in the above equations, we obtain
\begin{eqnarray}
\label{mod2}
s G(k_x,k_y,1;s) - P(k_x,k_y,1,t=0) & = & \gamma G(k_x,k_y,2;s) - \delta G(k_x,k_y,1;s),~~~~~~~ \mbox{ $l = 1$} \nonumber \\
s G(k_x,k_y,2;s) - P(k_x,k_y,2,t=0) & = & A(k_x,k_y) G(k_x,k_y,2;s)+ \delta G(k_x,k_y,1;s)+  \nonumber \\
                                    &   & \gamma G(k_x,k_y,3;s)-2 \gamma G(k_x,k_y,2;s),~~~~~ \mbox{ $l = 2$} \nonumber \\
s G(k_x,k_y,l;s) - P(k_x,k_y,l,t=0) & = & A(k_x,k_y) G(k_x,k_y,l;s)+ \nonumber \\
                                    &   & \gamma [G(k_x,k_y,l-1;s)+ \nonumber \\
                                    &   &  G(k_x,k_y,l+1;s)-2 G(k_x,k_y,l;s)],~~~ \mbox{ $l \geq 3$}
\end{eqnarray}
We have used the following definitions
\begin{eqnarray}
\label{def1}
 G(k_x,k_y,l;s | 0,0,l_0;t_0) & = & \int_{0}^{\infty} e^{- s t}
              \sum_{n,m=-\infty}^{\infty} e^{k_x n + k_y m}
               P(n,m,l;t | n,m,l_0;t_0) dt  \nonumber \\
                             & = & \mbox{} L[\sum_{n,m=-\infty}^{\infty} e^{k_x n + k_y m} P(n,m,l;t | n,m,l_0;t_0)],
\end{eqnarray}
where $L$ indicates the Laplace transform of the
quantity within the brackets, and
\begin{equation}
\label{def2}
 A(k_x,k_y) = 2 \alpha [Cos(k_x)-1] +
           2 \beta [Cos(k_y)-1].
\end{equation}
It is possible to write Eq. (\ref{mod2}) in  matrix form as
\begin{equation}
\label{mat1}
 [u \tilde{I} - \tilde{H}] \tilde{G} = \delta_{l l_0}= \tilde{I}_{l l_0},
\end{equation}
where the square matrix $\tilde{G}$ has components
\begin{equation}
\label{mat2}
\tilde{G}_{l l_0} = [G[k_x,k_y,l;s | n,m,l_0;t_0 ]],
\end{equation}
In Eq. (\ref{mat1}), $\tilde{I}$ is the identity matrix and
$\tilde{H}$ is a three-diagonal matrix with the following form
\[\tilde{H}=\left( \begin{array}{cccccc}
-\delta &  \gamma       & 0       & 0      & 0        & \cdots  \\
 \delta &  C            & \gamma  & 0      & 0        & \cdots  \\
 0      &  \gamma       & C       & \gamma & 0        & \cdots   \\
 0      &  0            & \gamma  & C      & \gamma   & \cdots  \\
 \cdot  &  \cdot        & \cdot   & \cdot  & \cdot    & \cdots   \\
                   \end{array}  \right), \]
and $C$ is defined as
\begin{equation}
\label{C}
C = -2 \gamma + A(k_x,k_y).
\end{equation}
In order to find the solution to the Eq. (\ref{mat1}), we
decompose the $\tilde{H}$ matrix in the following way
\begin{equation}
\label{mat4}
\tilde{H}=A(k_x,k_y) \tilde{I} +\tilde{H}_0 + \tilde{H}_1 + \tilde{H}_2,
\end{equation}
where
\[\tilde{H}_0=\left( \begin{array}{cccccc}
-2 \gamma & \gamma       & 0         & 0         & 0      & \cdots \\
 \gamma   &  -2 \gamma   & \gamma    & 0         & 0      & \cdots \\
 0        &  \gamma      & -2 \gamma & \gamma    & 0      & \cdots \\
 0        &  0           & \gamma    & -2 \gamma & \gamma & \cdots \\
 \cdot    &  \cdot       & \cdot     & \cdot     & \cdot  & \cdots \\
                   \end{array}
              \right), \]

\[(\tilde{H}_1)_{i j}=\Delta_1 \left\{ \begin {array}{ll}
                           1      & \mbox{if i = j = 1} \\
                           0      & \mbox{otherwise}
                         \end{array}
                \right. \]

\[(\tilde{H}_2)_{i j}=\Delta_2 \left\{ \begin {array}{ll}
                           1      & \mbox{if i = 1 and j = 2} \\
                           0      & \mbox{otherwise}
                         \end{array}
                \right. \]
with
\begin{eqnarray}
\label{deltas}
\Delta_1 & = & -\delta - [-2 \gamma + A(k_x,k_y)], \nonumber \\
\Delta_2 & = & \delta - \gamma.
\end{eqnarray}
Defining
\begin{eqnarray}
\label{lasG}
\tilde{G}^0 & = & [(s - (A(k_x,k_y)) \tilde{I} - \tilde{H}_0)]^{-1}, \nonumber \\
\tilde{G}^1 & = & [(s - (A(k_x,k_y)) \tilde{I} - \tilde{H}_0- \tilde{H}_1)]^{-1},
\end{eqnarray}
observing that a formal solution of Eq. (\ref{mat1}) is
\begin{equation}
\label{sol}
 \tilde{G} =  [s \tilde{I}-\tilde{H}]^{-1},
\end{equation}
and by reiterating the Dyson formula, we can show that
\begin{equation}
\label{G2}
 \tilde{G}_{l l_0} = \tilde{G}^{1}_{l l_0}+ \frac{\Delta_2 \tilde{G}^{1}_{l 2} \tilde{G}^{1}_{1 l_0}}
                           {1-\Delta_2 \tilde{G}^{1}_{1 2}},
\end{equation}

\begin{equation}
\label{G1}
 \tilde{G}^{1}_{l l_0} =  \tilde{G}^{0}_{l l_0} +
                          \frac{\Delta_1 \tilde{G}^{0}_{l 1} \tilde{G}^{0}_{1 l_0}}
                           {1-\Delta_1 \tilde{G}^{0}_{1 1}}.
\end{equation}

The form for $ \tilde{G}^{0}_{l l_0}$ can be obtained by
conventional methods \cite{vanKampen} as
\begin{equation}
\label{G0}
 \tilde{G}^{0}_{l l_0} =  L(e^{2 \gamma t} (I_{|j-j_0|}(2 \gamma t) -
                          I_{|j+j_0|}(2 \gamma t)))_{\tilde{u}},
\end{equation}
where $I_{n}(2 \gamma t)$ is the first modified Bessel function of
order $n$. The above expression points out that the Laplace
transform is evaluated at the argument $\tilde{u}=[s-A(k_x,k_y)]$.

Once the general expression for $\tilde{G}$ is obtained, we can
find the probability that a particle is on the plane at site
$(m,n,l = 1)$ at time $t$ given it was at $(0,0,l=1)$ at $t=0$.
This probability may be obtained using the inverse Laplace
transform in $s$ and the inverse Fourier transform on $k_x$, $k_y$
of the matrix element $\tilde{G}_{11}$.

A direct measurable experimental magnitude \cite{Bichuk} is the
variance of the probability distribution at time $t$ over the
plane  $z=1$
\begin{equation}
\label{variance}
  <r^2(t)>_{plane},
\end{equation}
which measures the spreading of particles over this plane. Once
$P(m,n,l=1; t | 0,0,l_0=1;t=0)$ is known, the variance is
calculated as:
\begin{equation}
\label{variance2}
<r^2(t)>_{plane} = \sum_{m,n=-\infty}^{\infty} P(m,n,l=1; t | 0,0,l_0=1;t_0) (m^2+n^2).
\end{equation}
Here, we have used symmetry properties for the diffusion along the
$x$ and $y$ axes, that is $<x(t)> = <y(t)> = 0$.

The Laplace transform of the variance $<r^2(s)>_{plane} =
L(<r^2(t)>_{plane})$ can be found as follows
\begin{equation}
\label{variance3} <r^2(s)>_{plane} = -
[\frac{\partial^{2}}{\partial
k_{x}^{2}}+\frac{\partial^{2}}{\partial k_{y}^{2}}]
[\tilde{G}_{11}]\mid_{k_x=k_y=0}
\end{equation}
By using Eqs. (\ref{sol}) to (\ref{variance3}), $<r^2(u)>_{plane}$
turns out to be the ratio of two complicated functions of $s$
\begin{equation}
\label{variance4}
<r^2(s)>_{plane} =\frac{N(s)}{D(s)},
\end{equation}
where
\begin{equation}
\label{num}
N(s) = [4 \delta \gamma (\alpha + \beta)] [s^2(\sqrt{s (4 \gamma + s)}
-3 s) + 3 \gamma^2 (\sqrt{s (4 \gamma + s)}- 3 s) - 2 \gamma^3],
\end{equation}
and
\begin{equation}
\label{den}
D(s) = \sqrt{s (4 \gamma + s)} (\gamma s (2 \gamma + s
- \sqrt{s (4 \gamma + s)}) + \delta (\gamma (\sqrt{s (4 \gamma +
s)}- 3 s) + s (\sqrt{s (4 \gamma + s)}-s)))^2.
\end{equation}
It is  important to remark that the conservation of particles
in the plane is not satisfied.

If we denote with $P_{plane}(t)$ the probability that the particle
is  in the plane $z=1$ at time $t$, it can be shown that the
Laplace transform of the magnitude is
\begin{equation}
\label{LaplaceP}
\tilde{P}_{plane}(s)=
\frac{(\gamma (2 \gamma + s - \sqrt{s (4 \gamma + s)})}
     {(\gamma s (2 \gamma + s - \sqrt{s (4 \gamma + s)})+\delta
      (\gamma (\sqrt{s (4 \gamma + s)}- 3 s)+ s (\sqrt{s (4 \gamma + s)}-s)))}.
\end{equation}

\begin{figure}
\centering
\resizebox{.6\columnwidth}{!}{\includegraphics{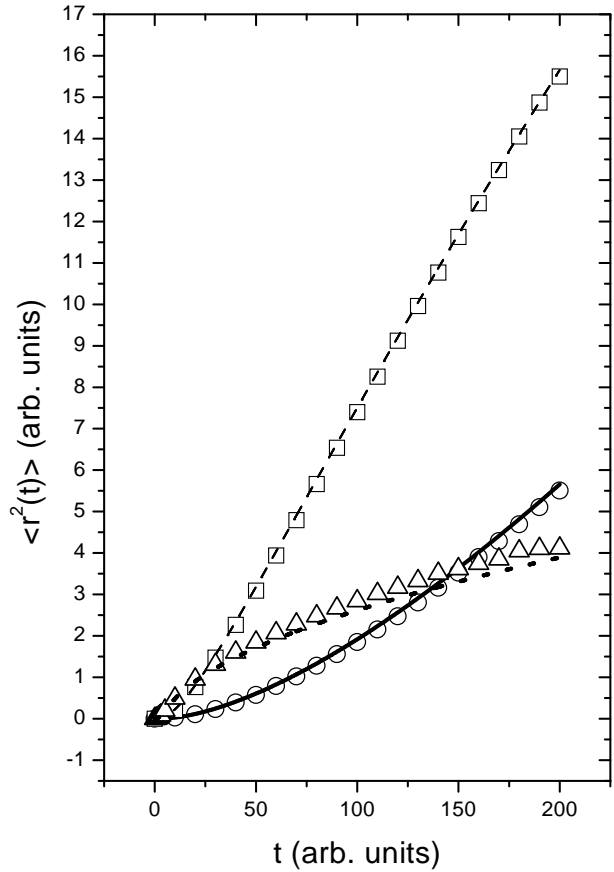}}
\caption{$<r^2(t)>$ vs time, for three different values of
$\delta$. Circles correspond to simulations with $\delta = 0.01$,
while the solid line is the related theoretical curve. Squares are
for simulations with $\delta = 0.1$, and the dashed line is the
theoretical curve. Triangles are for simulations with $\delta =
1.0$, and the dotted line is the corresponding theoretical curve.
In all figures we have used $\alpha = \beta = \gamma = 1.0$, and
the number of realizations is $10^{6}$.} \label{f25}
\end{figure}

\begin{figure}
\centering
\resizebox{.6\columnwidth}{!}{\includegraphics{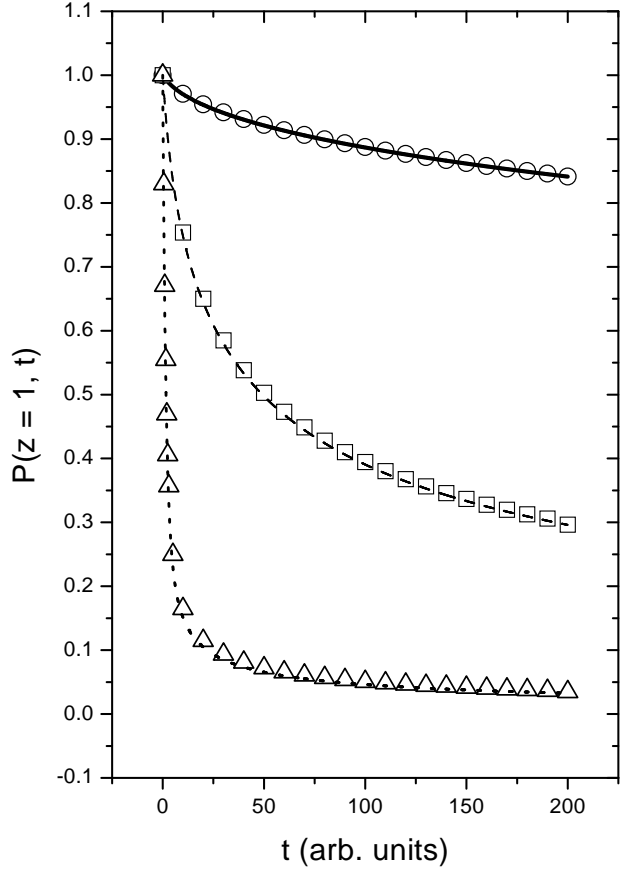}}
\caption{Probability $P(z=1;t)$ vs time, for three different
values of $\delta$. Circles correspond to simulations with $\delta
= 0.01$, while the solid line is the theoretical curve. Squares
are for simulations with $\delta = 0.1$, and the dashed line is the
theoretical curve. Triangles are for simulations with $\delta =
1.0$, and the dotted line is the corresponding theoretical curve.}
\label{f15}
\end{figure}

In order to test the theoretical results for the
$<r^2(t)>_{plane}$ and $P_{plane}(t)$, we have performed
Montecarlo simulations for the adsorption-desorption processes
obtaining an excellent agreement in both cases. Figures  \ref{f25}
and \ref{f15} show the variance and the probability that the
particle is on the plane $z=1$ as a function of $t$, for three
different values of $\delta$.

\begin{figure}
\centering
\resizebox{.6\columnwidth}{!}{\includegraphics{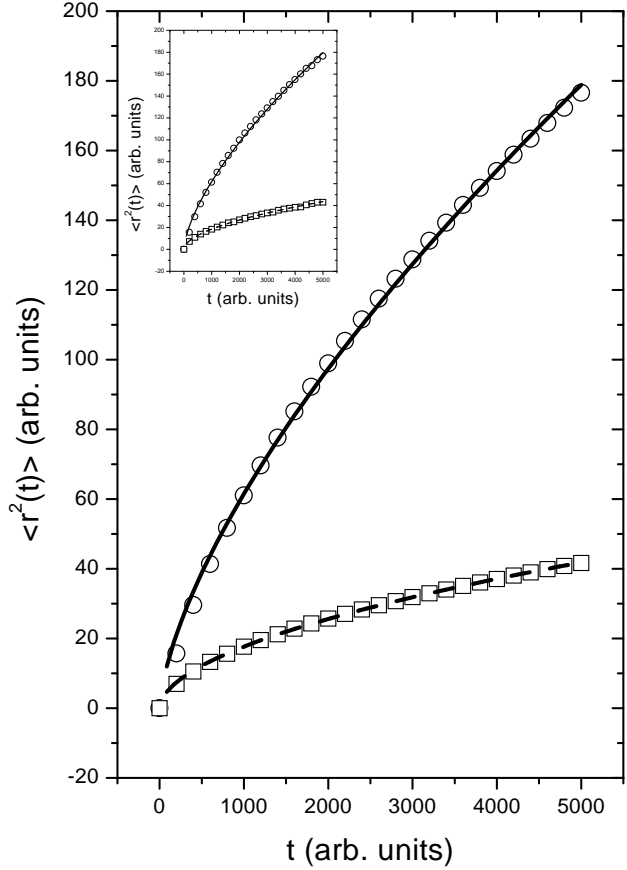}}
\caption{$<r^2(t)>$ vs time, for three different values of
$\delta$, and compared with a fit using $y=a  t^{
\epsilon }$. Circles correspond to theory with $\delta = 0.01$,
and the solid line to the fit with $a=0.63\pm 0.03$,
$\epsilon=0.663\pm 0.007$. Squares are for theory with $\delta =
0.1$, while the solid line is the fit with $a=0.435 \pm
0.007$, $\epsilon=0.535 \pm 0.005$. The insert shows the related
results for simulations. Again circles correspond to $\delta =
0.01$, and the solid line to the fit with $a=0.64\pm 0.04$ and
$\epsilon=0.661\pm 0.008$; while squares are for $\delta = 0.1$,
and the solid line is the fit with $a=0.47 \pm 0.01$,
$\epsilon=0.531 \pm 0.004$. } \label{f35}
\end{figure}

The asymptotic behavior for large $t$ of both, $<r^2(t)>_{plane}$
and $P_{plane}(t)$, can be obtained by means of Tauberian theorems
\cite{Montroll} as
\begin{equation}
\label{asimpr}
<r^2(t)>_{plane} \rightarrow \frac{\sqrt{\gamma}}{\delta} \,\,\,\,\
\frac{(\alpha + \beta)}{\Gamma[3/2]} \,\,\,\,\ t^{\frac{1}{2}},
\end{equation}
\begin{equation}
\label{asimpP}
P_{plane}(t) \rightarrow  \frac{\sqrt{\gamma}}{\delta} \,\,\,\,\
\frac{(\alpha + \beta)}{\Gamma[1/2]} \,\,\,\,\ t^{-\frac{1}{2}}.
\end{equation}
From Eq. (\ref{asimpr}) we recognize an asymptotic sub-diffusive
regime which is shown in Fig. \ref{f35}.

\begin{figure}
\centering
\resizebox{.6\columnwidth}{!}{\includegraphics{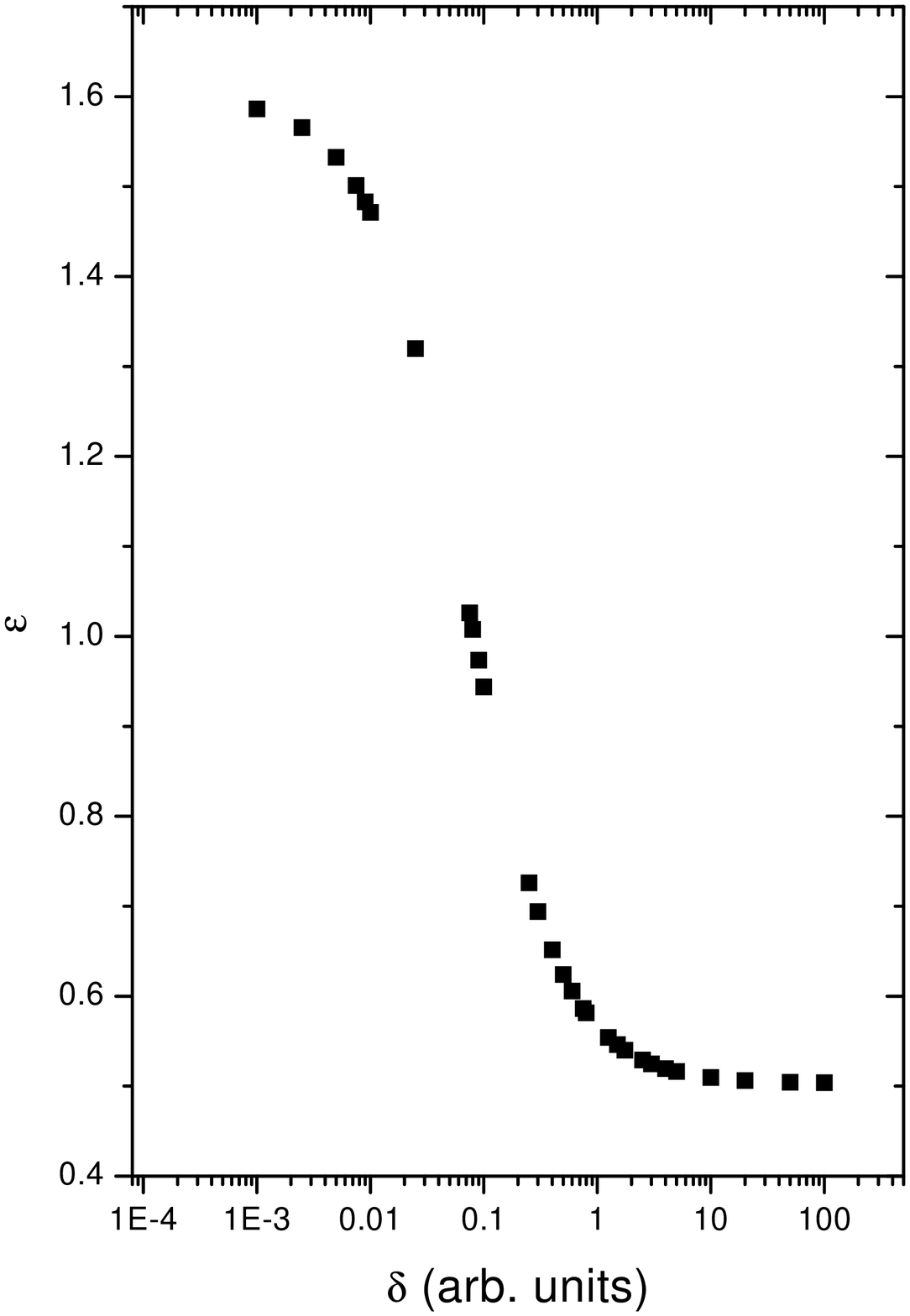}}
\caption{$\epsilon $ vs $\delta $, for  time in the range $t = 0$
to $t= 500$. Note that for $\delta \rightarrow \infty$, $\epsilon$
approaches the value $0.5$, while for small values of $\delta $,
$\epsilon$ becomes $\gg 1$. } \label{f55}
\end{figure}

When we choose a different range of time in order to fit
$<r^2(t)>_{plane}$ as $t^{\epsilon}$ we find that $\epsilon$
depends on the values of $\delta$ for fixed $\alpha, \beta$ and
$\gamma $. Figure \ref{f55} shows this dependence
for a wide range of values of the desorption rate $\delta$.

\section{The CTRW Scheme}
\label{ctrw}

In this section we sumarize the most important results of the
lattice {\it CTRW} approach. Let $\Psi(\vec{r},\vec{r'};t)$ the
waiting time density governing a single transition being defined
as the probability density that a transition from $\vec{r'}$ to
$\vec{r}$ occurs at time between $t$ and $t+ dt$ given the walker
arrives at $\vec{r}$ at $t=0$. The transition time between
different sites is assumed much shorter than the time spent at
each lattice point; these "instantaneous transitions"  allow us to
prove probability conservation as indicated below.

We assumed translational invariance, that is $\Psi(\vec{r},
\vec{r'}; t) = \Psi(\vec{r}-\vec{r'};t)$ for all $\vec{r}$ and
$\vec{r'}$, hence the conditional probability to find the particle
at site $\vec{r}$ at time $t$ given it was at $\vec{r}=0$ at
$t=0$, $P(\vec{r},t | \vec{r}=0,t=0)$ can be found in
Laplace-Fourier variables $\vec{k}$ and $s$ as
\begin{equation}
\label{ctrw1} P(\vec{k},s | \vec{r}=\vec{0};t=0) =
\frac{1}{1-\tilde{\Psi}(\vec{k},s)}
                   \,\, \frac{1-\tilde{\Psi}(s)}{s}
\end{equation}
where $\tilde{\Psi}(\vec{k},s)$ is the Fourier-Laplace transform
of $\Psi(\vec{r},t)$ and $\tilde{\Psi}(s) =
\Psi(\vec{k}=\vec{0},s)$ is the Laplace transform of the total
transition probability.

The normalization of $P(\vec{r},t | \vec{r}=\vec{0},t=0)$ for all
$t$ is obtained from Eq. (\ref{ctrw1}) in a direct way as
\begin{equation}
\label{ctrw2} \sum_{\vec{r}}P(\vec{r}; t | \vec{r}=\vec{0};t=0) =
L^{-1}(\tilde{P}(\vec{k}=\vec{0}, s | \vec{r}=\vec{0},t=0)) =
L^{-1}(\frac{1}{s}) = 1.
\end{equation}
Normalization implies that the walker is somewhere in the (3D)
lattice, but if we evaluate the probability that the walker is in
a subspace of this lattice we will lose the normalization
condition.

We now consider, as done previously \cite{Bichuk}, the possibility
that the successive visits of the walker to the plane $z = 1$, as
discussed in Section \ref{modelo}, may be viewed as an CTRW over
that plane.  We must define a suitable waiting time density for
single transitions, $\Psi_{plane}(\vec{r},\vec{r'};t)$ where
$\vec{r}$ and $\vec{r'}$ are two dimensional vectors on the plane,
i.e. $\vec{r} = (m,n)$ and $\vec{r'} = (m', n')$. A transition
between $\vec{r}$ and $\vec{r'}$ must be done with no visit to the
plane within the interval $(0, t)$, in order to be consistent with
a single transition. If we remember that this waiting time density
is the probability density that the walker arrives at $\vec{r}$
between $t$ and $t+dt$ having arrived at $\vec{r'}$ at $t = 0$, it
is obvious that this transition is not instantaneous because the
''flying time" across the bulk cannot be neglected (see below). In
order to be consistent with the CTRW theory we may consider that
the particle remains at site $\vec{r'}$ during this time and then
perform  a jump to the site  $\vec{r}$. In this way, the
probability over the plane is conserved and transitions become
instantaneous.

Now we build up the waiting time density for this "single
transition" in the plane $z = 1$, taking into account the above
remarks. We note that if the walker has arrived at $(m', n', 1)$
at $t = 0$, the probability density to desorb  from the plane per
unit time around $t'$ by a jump to $(m', n', 2)$ is $\,\sim \delta
\,\exp(- \delta t')$. We define $q[(m,n,2), t; (m',n',2),t']$ as
the probability of finding the walker at $(m,n,2)$
 at time $t$ given it was in $(m',n',2)$ at time $t'$
without visiting the plane $z=1$ in the interval $(t', t)$. The
probability density to reach $(m,n,1)$ for the first time between
$t$ and $t + dt$ given the walker was in $(m',n', 2)$ at time $t'$
can be expressed as
\begin{equation}
\label{ctrw3}
f[(m,n,1),t;(m',n',2),t')] = \gamma q[(m,n,2),t; (m',n',2),t'].
\end{equation}

Finally the  density  to reach $(m,n,1)$ for first time, per unit
of time around $t$ given that the walker was in $(m',n',1)$ at
$t=0$ without visiting the plane in the interval $(0,t)$ is:
\begin{equation}
\label{ctrw4}
f[(m,n,1),t; (m',n',1),0] = \int_0^t [\gamma q[(m,n,2),t; (m',n',2),t']]
                            \delta \,\exp(-\delta t') dt'.
\end{equation}
The function $f[(m,n,1),t; (m',n',1),t']$ plays the role of the
waiting time density $\Psi_{Plane}(\vec{r},\vec{r'};t)$ where
$\vec{r}=(m,n)$ and $\vec{r'}=(m',n')$.
Since we are assuming  translational invariance in the $x$ and $y$
directions, the function $q$  depends only on $(m-m')$ and
$(n-n')$. Selecting $(0,0)$ as the starting point we  obtain
\begin{equation}
\label{ctrw5} \Psi_{plane}(m,n,t) = \int_0^t [\gamma q(m,n,2),t;
(0,0,2),t'] \delta \,\exp(-\delta t') dt'.
\end{equation}
The function $q((m,n,2),t; (0,0,2),t')$ is obtained by means of
the method of images assuming an absorbent plane in $z = 1$
\begin{eqnarray}
\label{ctrw6}
q(m,n,2), t; (0,0,2), t') & = & [\exp(- 2 \alpha t) I_m(2 \alpha t)] \nonumber \\
                          &   & [\exp(- 2 \beta t) I_n(2 \beta t)] \nonumber \\
                          &   & [\exp(-2 \gamma t) I_0(2 \gamma t) - \exp(-2 \gamma t I_2(2 \gamma t)],
\end{eqnarray}
where $I_j$ is the modified Bessel function of order $j$.
Equations (\ref{ctrw5}) and (\ref{ctrw6}) allow us to build the
''normalized probability" of the CTRW in the plane $z = 1$
\begin{equation}
\label{ctrw7}
P^{norm}(\vec{r},t | \vec{r}=\vec{0}, t=0) = L^{-1} F^{-1} \left(\frac{1}{1-\Psi_{plane}(\vec{k},s)}\right)
                                             \left(\frac{1-\Psi_{plane}(s)}{s}\right),
\end{equation}
where
\begin{equation}
\label{ctrw8}
\Psi_{plane}(\vec{k},s) = \gamma (\frac{1}{2 \gamma^2}) (s + 2 \gamma + A(k_x,k_y) -
                           \sqrt{(s+2 \gamma -A(k_x,k_y)^2-(2 \gamma)^2})) \frac{\delta}{s+\delta},
\end{equation}
and $ \Psi_{Plane}(s)=\Psi_{Plane}(\vec{k}=\vec{0},s)$. Here
$\vec{r} = (m,n); \vec{k}=(k_x,k_y)$ and the function $A(k_x,k_y)$
is defined by Eq. (\ref{def2}). Eq. (\ref{ctrw8}) shows a
''coupled" waiting time density with a divergent time first
moment, that is $<t> \rightarrow \infty $.

Expressions for the Fourier-Laplace transform of
$P^{norm}(\vec{r},t | \vec{r}=\vec{0}, t=0)$ can be obtained from
Eqs. (\ref{ctrw7}) and (\ref{ctrw8}).
\begin{equation}
\label{ctrw9} P^{norm}(\vec{k},s | \vec{r}=\vec{0}, t=0) =
\frac{\frac{1}{s} (2 \gamma (s+\delta) - \delta (s + 2 \gamma -
\sqrt{s (s+ 4 \gamma)}))} {2 \gamma (s+\delta)-\delta (s + 2
\gamma + A(k_x,k_y) - \sqrt{(s+2 \gamma + A(k_x,k_y))^2-(2
\gamma)^2})}.
\end{equation}
The corresponding variance of $P^{norm}$ in Laplace space is
\begin{equation}
\label{ctrw10} <r^2(s)>^{norm} = 2 \delta (\alpha + \beta)
(\frac{2 \gamma + s - \sqrt{s (s+ 4 \gamma)}} {2 \gamma s + \delta
\sqrt{s (s+ 4 \gamma)}-s} s \sqrt{s (s+ 4 \gamma)}),
\end{equation}
and the asymptotic behavior for large $t$ is
\begin{equation}
\label{ctrw10p} <r^2(t)>^{norm} \rightarrow (\alpha + \beta) \, t.
\end{equation}
This result, i.e. normal diffusion, is due to the coupled
character of the waiting time density Eq. (\ref{ctrw8}), and its
infinite time first moment.

\begin{figure}
\centering
\resizebox{.6\columnwidth}{!}{\includegraphics{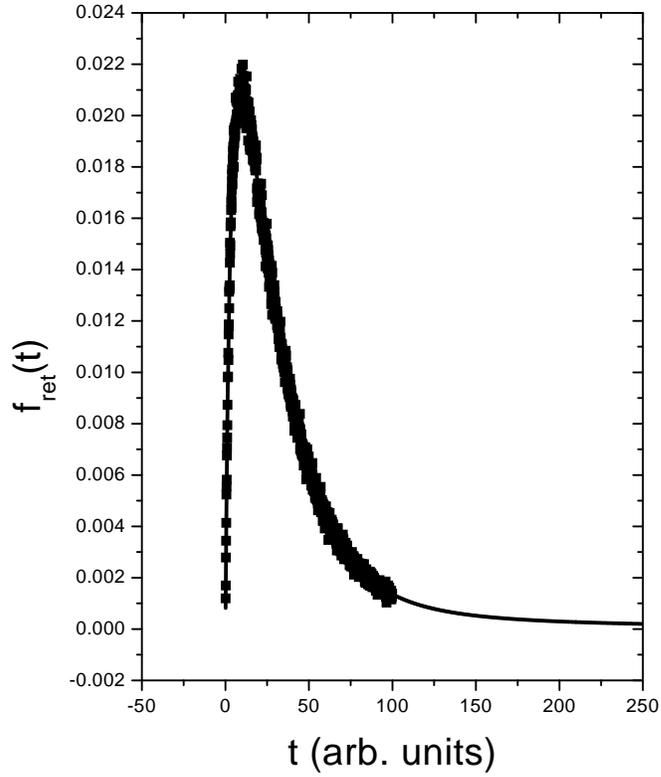}}
\caption{Distribution function for return times $f_{ret}(t)$ vs
time for $\delta = 0.5$. The squares correspond to simulations
while the solid line is for theory. } \label{f65}
\end{figure}

As an important byproduct of the above approach, we present the
evaluation of the probability of the first return to the plane $z
= 1$. If a walker, initially at the point $(0,0,1)$, desorbs and
begins an excursion across the bulk, the probability density to
return, for the first time, to the plane $z = 1$  between $t$ and
$t + dt$, $f_{ret}(t)$, is given by
\begin{equation}
\label{ctrw11} f_{ret}(t) = L^{-1}[\Psi_{plane}(s)] =
L^{-1}[\Psi_{plane}(\vec{k}=\vec{0},s)].
\end{equation}
From the Eq. (\ref{ctrw8}) we obtain the following expression the
Laplace transform of the first return density
\begin{equation}
\label{ctrw12}
\Psi_{plane}(s) = (\frac{\delta}{2 \gamma}) (\frac{1}{s + \delta})
                  (2 \gamma + s - \sqrt{u (s + 4 \gamma)}).
\end{equation}
We have made a numerical inverse transform of this result by using
a numerical program and have compared this result with Montecarlo
simulations finding an excellent agreement. See Fig.  \ref{f65}.

Finally, from Eq. (\ref{ctrw12}) it is possible to obtain two
important results. Firstly, a walker is certain to return to the
plane
\begin{equation}
\label{ctrw13} \int_{0}^{\infty} f_{ret}(t) \, dt =
\Psi_{plane}(s=0) = 1.
\end{equation}
Secondly, the asymptotic (long time) behavior of the first return
density is
\begin{equation}
\label{ctrw14} f_{ret}(t) \rightarrow
(\frac{1}{\sqrt{\gamma}\,\,\,\ \Gamma[\frac{1}{2}]})
\frac{1}{t^{\frac{3}{2}}},
\end{equation}
indicating that the mean first time to return to the plane is
infinite.

\section{Conclusions}

We presented in this paper an analytical model for the
adsorption-desorption processes from a boundary plane in a
semi-infinite cubic lattice. We  studied the  effective diffusion
of  molecules on this plane interface and calculated  the
evolution of the square width of the probability distribution.
This square width can be fitted with a power law of $t$ whose
exponent changes with the range of time considered  and depends on
the values of the adsorption and diffusion parameters. In this
sense, effective anomalous super-diffusions  reported by Bichuk
and O$´$Shaughnessy \cite{Bichuk} may be understood. However, in
the asymptotic long time regime this square width always behaves
as $t^{\frac{1}{2}}$. It is important to observe that the lack of
probability conservation over the  interface must be taken into
account if a "genuine"  CTRW on the plane is considered as was
pointed out in Section \ref{ctrw}. As a byproduct of our approach,
we obtained the evaluation of the probability of the first return
to the planar interface. We performed Montecarlo simulations of
the adsorption-desorption process obtaining excellent agreement
with the model predictions.

This work is part of a research project on bulk mediated diffusion
on surfaces. Here we have discussed the case of infinite bulk
while in \cite{nosotros1} we investigated the finite (in the
direction normal to the surface) case where, among other aspects,
we have found that an ``optimal" number of layers exists, that
produces the faster growth of $<r^2(t)>$. In addition, we have
also investigated the case of non-Markovian desorption process
\cite{nosotros2} where we have found an interesting oscillatory
behavior. This research offers a more or less complete view of the
theoretical description for the problem of bulk mediated diffusion
on a surface.

\vspace{0.25cm}

{\bf Acknowledgments:} The authors thank V. Gr\"unfeld for a
critical reading of the manuscript. HSW acknowledges the partial
support from ANPCyT, Argentine, and thanks the MECyD, Spain, for
an award within the {\it Sabbatical Program for Visiting
Professors}, and to the Universitat de les Illes Balears for the
kind hospitality extended to him.


\begin{thebibliography}{99}



\bibitem{v1} J. H. Clint, {\it Surfactant Aggregation}
(Chapman and Hall, New Jork, 1992).

\bibitem{v2} H. E. Johnson, J. F. Douglas and S. Granick,
Phys. Rev. Lett.   {\bf 70} 3267 (1993).

\bibitem{v3} C.T. Shibata and A. M Lenhof, {\it J. Colloid
Interface} Sci. {\bf 148}, 469 (1992), {\bf 148}, 485 (1992).

\bibitem{v4} S. Kim and H. Y, J. Phys. Chem.  {\bf 96}, 4034
(1992).

\bibitem{v5} Y. L. Chen, S. Chen, C. Frank and J. Israelachvili,
{\it J. Colloid Interface Sci.} {\bf 153}, 244  (1992).

\bibitem{v6} A. A. Sonin, A. Bonfillon, and D. Langevin,
Phys. Rev. Lett. {\bf 71}, 2342 (1993).

\bibitem{v7}A. L. Adams, G. C. Fishe, P. C. Munoz and L. Vroman,
J. Biomed. Meter. Res. {\bf 18}, 643 (1984).

\bibitem{Bichuk} O. Bichuk and B. O´Shaughnessy, J. Chem. Phys.,
{\bf 101}, 772 (1994). O. Bichuk and B. O´Shaughnessy, Phys. Rev.
Lett., {\bf 74}, 1795 (1995). S. Stapf, R. Kimmich and R. O.
Seitter, Phys. Rev. Lett., {\bf 75}, 2855 (1995).

\bibitem{Tsallis} C. Tsallis, S.V.F. Levy, A.M.C. Souza and R.
Maynard, Phys. Rev. Lett. {\bf 75}, 3598 (1995). [Erratum: Phys.
Rev. Lett. {\bf 27}, 5442 (1996)].

\bibitem{Prato} D. Prato and C. Tsallis, Phys. Rev. E. {\bf 60},
2398 (1999).

\bibitem{Re} M. A. R\'e, C. E. Budde and D. P. Prato, Physica A,
{\bf 323}, 9 (2003).

\bibitem{Zasla} G. M. Zaslavsky, Physica D, {\bf 76}, 110 (1994).

\bibitem{Klafter} J. Klafter, A. Blumen and M.F. Shlesinger, Phys.
Rev. A {\bf 35}, 3081 (1987).

\bibitem{Blumen} A. Blumen, G. Zumofen and J. Klafter, Phys.Rev. A
{\bf 40}, 3964 (1989).

\bibitem{Zumofen} G. Zumofen, A. Blumen and M.F. Shlesinger, J.
Stat. Phys.{\bf 54}, 1519 (1889).

\bibitem{vanKampen} N.G. Van Kampen, {\it  Stochastic Processes in
Physics ans Chemistry}, (North-Holland, Amsterdam, 1981).

\bibitem{Montroll} E.W. Montroll and B.J. West, in {\it
Fluctuation Phenomena}, E.W. Montroll and J.L. Lebowitz, eds.
(North Holland, Amsterdam, 1979).

\bibitem{nosotros1} J.A. Revelli, C.E. Budde, D. Prato and H.S.
Wio, {\it Bulk Mediated Surface Diffusion: Finite System Case}, to
be submited.

\bibitem{nosotros2} J.A. Revelli, C.E. Budde, D. Prato and H.S.
Wio, {\it Bulk Mediated Surface Diffusion: Non Markovian
Dynamics}, to be submited.

\end{thebibliography}
\end{document}